# Electron excitation and energy transfer rates for $H_2O$ in the upper atmosphere

Penny Thorn, Laurence Campbell* and Michael Brunger

Address: ARC Centre for Antimatter-Matter Studies, SoCPES, Flinders University, GPO Box 2100, Adelaide, SA 5001, Australia

Email: Penny Thorn - penny.thorn@flinders.edu.au; Laurence Campbell* - laurence.campbell@flinders.edu.au; Michael Brunger - michael.brunger@flinders.edu.au

* Corresponding author

**Abstract**

Recent measurements of the cross sections for electronic state excitations in $H_2O$ have made it possible to calculate rates applicable to these excitation processes. We thus present here calculations of electron energy transfer rates for electronic and vibrational state excitations in $H_2O$, as well as rates for excitation of some of these states by atmospheric thermal and auroral secondary electrons. The calculation of these latter rates is an important first step towards our aim of including water into a statistical equilibrium model of the atmosphere under auroral conditions.

## 1. Introduction

Collisions with neutral molecules are an important process by which electrons in the atmosphere lose energy [1]. The electron energy transfer rate, $Q_i$, for a particular excitation process, $i$, is a measure of the rate at which electrons lose energy in a collision with a molecule causing the excitation $i$. Hence, the electron energy transfer rates highly influence electron temperature ($T_e$), which is in turn an important value in the study of atmospheric physics and chemistry. On the other hand the excitation rate $k_i$ for an inelastic process $i$ is a measure of the rate at which the particular state $i$ is excited by secondary electrons in the environment under investigation (e.g. auroral) [2]. This in particular is a crucial quantity in any attempt to model our ionosphere.

These values, $Q_i$ and $k_i$, were calculated with the eventual aim of incorporating investigations of $H_2O$ into a statistical equilibrium model of the ionosphere. The ultimate aim of this is to determine if the OH Meinel bands (identified in the infrared spectrum of the night sky by Meinel [3,4]) are caused at least in part by electron driven processes. The statistical equilibrium approach was initially developed by Cartwright [5], and was implemented previously to predict the absolute population density of vibrationally and electronically excited $N_2$ (Campbell *et al.* [6]), $O_2$ (Jones *et al.* [7]) and NO (Cartwright *et al.* [8]) over a range of altitudes. $H_2O$ will also ultimately be included in our ionospheric model in order to calculate population densities for OH, which is created by the dissociation of $H_2O$.

The Meinel bands are known to be caused by vibrationally excited OH molecules emitting photons, on decay towards the zero vibrational level of the ground electronic state. However the source of this vibrationally excited OH could be chemical or electron driven, or a mixture of both. The $\tilde{a}^3B_1$, $\tilde{A}^1B_1$, $\tilde{b}^3A_1$, and $\tilde{B}^1A_1$ excited states of $H_2O$ are all known to dissociate into OH, and hence excitation of these states may be an intermediate step towards the production of the Meinel bands. It is also believed [9] that other higher lying excited electronic states of $H_2O$ can dissociate into OH, by first de-exciting into one of those above four states. Such de-excitation in the atmosphere could be caused through the emission of a photon or by heavy-particle quenching.

A thorough study of electronic state excitation cross sections has made the calculation of these energy transfer and excitation rates possible for $H_2O$. Full details of those experiments can be found in Thorn *et al.* [10,11] and Brunger *et al.* [12], along with comparisons of the measurements with previous theoretical values. These were the first experimental studies to publish absolute values for the electronic state excitation cross sections in $H_2O$ and covered the incident electron energy range 15 to 50 eV, and excitation energies up to 12 eV. In addition another 19 "composite" electronic states, so called as some of the electronic states cannot be resolved from one another, are reported here for the first time. All these cross sections for electronic state excitation were used in the present calculations, along with cross sections for vibrational excitation previously determined by Yousfi and Benabdessadok [13].

## 2. Results and discussion
### *2.1 Method*
The excitations covered in this work were the (010) bending mode vibration, the combined (100)+(001) stretching vibrational modes and the 25 electronic state excitations as listed in table 1. Note that the designations for the electronic states given in table 1 are discussed in detail in Thorn [14] and references therein. Further note that the 19 original integral electronic state cross sections are also plotted in figure 1. In this work we note that the electron energy transfer rates were calculated for electron temperatures up to 12000 K, while the electron impact excitation rates were calculated for altitudes between 80 to 350 km under conditions of a medium strength (IBC II[+]) aurora [6].

**Table 1: Excited States of H$_2$O**

| State Designation | Excitation Energy (eV) | State Number |
|---|---|---|
| (010) | 0.198 | 1 |
| (100)+(001) | 0.453 | 2 |
| $\tilde{a}^3B_1$ | 7.14 | 3 |
| $\tilde{A}^1B_1$ | 7.49 | 4 |
| $^3A_2$ | 8.90 | 5 |
| $^1A_2$ | 9.20 | 6 |
| $\tilde{b}^3A_1$ | 9.46 | 7 |
| $\tilde{B}^1A_1$ | 9.73 | 8 |
| $\tilde{d}^3A_1$ | 9.82 | 9 |
| $\tilde{c}^3B_1 + \tilde{C}^1B_1$ | 9.98 | 10 |
| $\tilde{D}^1A_1$ | 10.12 | 11 |
| $\tilde{C}^1B_1(100) + ^3B_1$ | 10.35 | 12 |
| $^1B_1 + \tilde{D}^1A_1(100)$ | 10.55 | 13 |
| $^3A_2$ | 10.70 | 14 |
| $\tilde{D}^1A_1(110) + \tilde{C}^1B_1(200)$ | 10.77 | 15 |
| $^1A_2$ | 10.84 | 16 |
| $\tilde{e}^3B_1 + \tilde{E}^1B_1 + \tilde{D}^1A_1(200)$ | 10.97 | 17 |
| $^3B_2 + ^1B_2 + ^3B_2$ | 11.10 | 18 |
| $^3A_1 + ^1A_2 + ^1A_1 + ^3A_2 + ^1B_1 + ^3B_1$ | 11.23 | 19 |
| $^1B_1 + ^3B_1 + ^3A_1 + ^1A_1$ | 11.35 | 20 |
| $^1B_2 + ^3B_2 + ^3B_1 + ^1B_1$ | 11.50 | 21 |
| $^3A_2 + ^1A_2$ | 11.61 | 22 |
| $^3B_1$ | 11.68 | 23 |
| $^1B_1$ | 11.75 | 24 |
| $^3A_1 + ^1A_1$ | 11.80 | 25 |
| $^1A_1 + ^3B_1 + ^1B_1$ | 11.90 | 26 |
| $^1A_1$ | 12.06 | 27 |

Table of the excitation energies (eV) for the excited states in water that we have considered in our calculations. The listed state numbers are also relevant to the data in figure 1.

Our method for calculating the energy transfer rates was the same as that described by Jones *et al.* [15], in that case for vibrational excitations in O$_2$. The electron energy transfer rate $Q_i$ for excitation from the zero vibrational level of the ground electronic state to electronic or vibrational level *i*, is given below in equation (1) [16]:

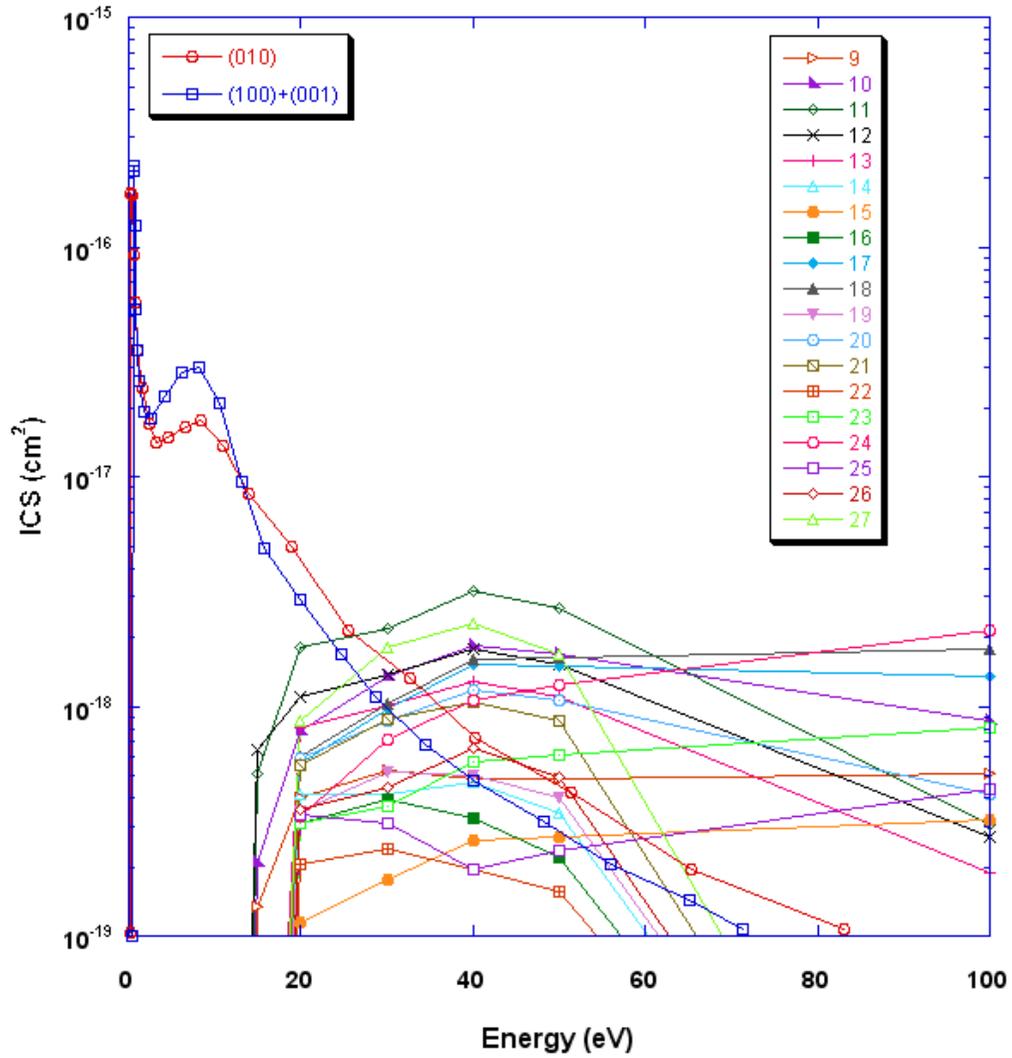

**Figure 1**
**Integral cross sections**. A selection of our linearly interpolated and extrapolated ICSs (cm$^2$), as a function of electron energy (eV), as used in the present computations. The electronic state excitation cross sections are original to this work, while the vibrational excitation cross sections are from Yousfi and Benabdessadok [13]. The respective excitation processes are as labelled on the figure.

$$Q_i = E_i \left( \frac{8k_B T_e}{p m_e} \right)^{\frac{1}{2}} \int_0^\infty s_i(x) x \exp(-x) dx, \quad (1)$$

where

$$x = \frac{E}{k_B T_e}.$$

In equation (1) $\sigma_i(x)$ is the integral cross section (ICS) for excitation from the zero vibrational level of the ground electronic state to the excited state $i$ (the cross sections employed in the present calculations are discussed in more detail later), $k_B$ is Boltzman's constant, $m_e$ is the electron mass and $E_i$ is the excitation energy for the transition of interest, as listed in table 1.

The electron impact excitation rates $k_i$, for the various excitations $i$, were computed using the equation [8]:

$$k_i = \int_0^\infty F(E)\sigma_i(E)dE. \qquad (2)$$

In equation (2) $\sigma_i(E)$ is again the integral cross section for the transition of interest, and we note that we used the same cross sections here that were employed in the calculation of the energy transfer rates. In this case $E$ is the electron energy and $F(E)$ is the electron flux at energy $E$. The electron flux spectra employed in the present calculations are discussed in more detail later.

### 2.1.1 Excitation cross sections

As mentioned previously the cross sections measured by Thorn and colleagues (e.g. [10-12]) were employed in calculating the energy transfer rates and excitation rates for the various electronic state excitations. We briefly note here that these cross sections were determined from electron energy loss spectra (EELS) measured with a crossed-beam electron spectrometer, at energies 15 to 50 eV and angles 10° to 90°. Differential cross sections (DCSs) were initially determined from the EELS using the elastic cross sections from Cho *et al.* [17] to put the values on an absolute scale. ICSs were then determined from the DCSs using a molecular phase shift analysis (MPSA) technique [18]. The overall uncertainty in the ICS measurements was approximately in the range 35% – 50% across the energy range considered. The cross sections for the vibrational excitations used in the present calculations were those of Yousfi and Benabdessadok [13], who employed an electron swarm parameter unfolding technique. These were used rather than the available ICS measurements [19] as they covered a larger energy range. Also, by definition, they were validated by being consistent with transport parameters and so considered to be more reliable. For (010) there is little difference between the swarm and ICS values below 6 eV, while for the (100)+(001) modes the swarm values are about half of the ICS values. However, these differences are of little consequence because the vibrational excitation will be dominated by the contribution from below 1 eV, where both the electron flux (see below) and the cross section are much larger.

In order to evaluate the integrals in equations (1) and (2), the cross sections employed in our calculations are each linearly interpolated and extrapolated beyond the final data point (50 eV for the electronic states and ≈ 80 eV for the vibrational cross sections). They are also extrapolated to a value of 0 at their respective threshold energies. Our previously reported cross sections can be found in references [10-12], with the new cross section data being reported here plotted in figure 1.

*2.1.2 Electron flux spectrum*

The electron flux spectra used in the present electron impact excitation rate calculations were the same as those detailed by Jones *et al.* [7] and Campbell *et al.* [6]. As a consequence we do not repeat that detail here except to note that these electron distribution spectra were dependent on altitude, and were a sum of the thermal electron distribution and that for the auroral secondary electron spectrum of interest to us here. The thermal electron spectrum was assumed to be a Maxwell-Boltzmann electron distribution, with the electron temperature and density taken from the International Reference Ionosphere [20,6]. The auroral secondary electron spectrum was determined using previous measurements from medium strength (IBC II[+]) aurorae as described by Campbell *et al.* [6]. However, using this method the minimum altitude of the resulting electron distribution was only 130 km. Since we wanted our excitation rate calculations to be made in the altitude range 80 to 350 km, the procedure described by Jones *et al.* [7] was used to extrapolate these data to lower altitudes. This method of extrapolation involved evaluating equation (3), for each altitude, $h$:

$$F_h = F_{120} e^{-0.1(120-h)}, \tag{3}$$

where,

$$F_{120} = F_{130} + \frac{E_{120} - E_{130}}{E_{350} - E_{130}} (F_{350} - F_{130}), \tag{4}$$

and

$$E_{h'} = 1 - e^{-0.027(h'-60)}. \tag{5}$$

In equations (3) to (5), $h'$ refers to any altitude, while $F_h$ is the secondary electron distribution at altitude $h$. $F_{130}$ and $F_{350}$ were determined using the shape of the electron distributions published by Lummerzheim and Lilensten [21], at 150 km and 300 km respectively, and then multiplying these distributions by a factor of 27 in order to obtain spectra of the same magnitude as those estimated by Feldman and Doering [22] and measured by Lummerzheim *et al.* [23] for an IBC II[+] aurora. Full details of this process can be found in Campbell *et al.* [6] and references therein. Examples of the resulting secondary electron distributions are plotted in figure 2, for 80 km, 130 km and 350 km altitudes.

*2.2 Results*

The 19 new integral cross sections for electron impact excitation of electronic states in $H_2O$ are plotted in figure 1, with no evidence for any resonance phenomena being noted. However, this is not surprising given the rather coarse energy grid of these measurements. Results from our cal-

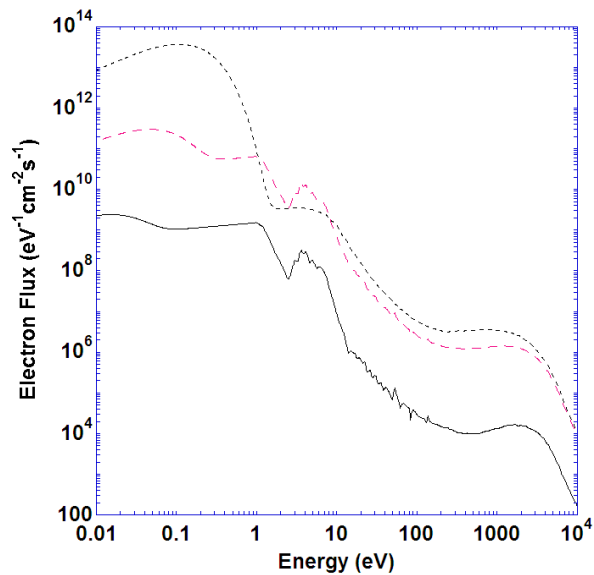

#### Figure 2
**Electron flux spectrum**. Electron flux (eV$^{-1}$cm$^{-2}$s$^{-1}$) as a function of electron energy (eV), as used in our calculation of the excitation rates. The respective altitudes for each secondary electron distribution are: (⎯) 80 km, (⎯ ⎯ ⎯) 130 km, (- - -) 350 km.

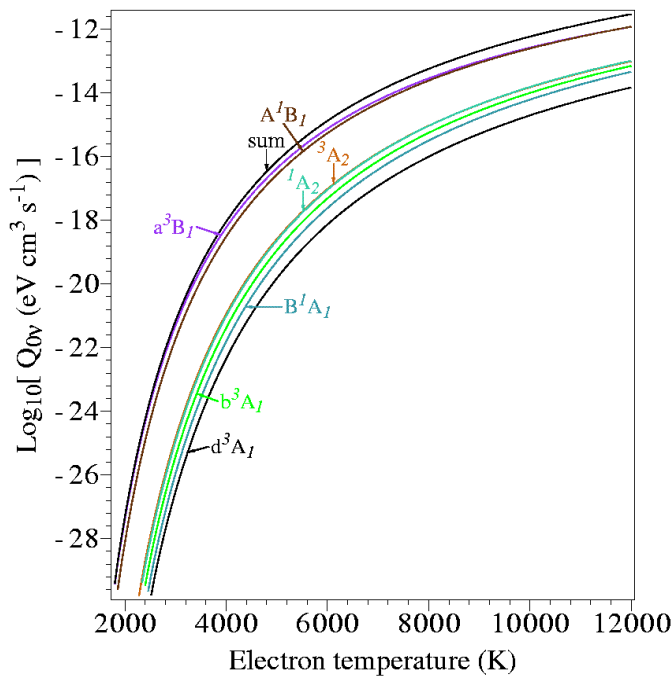

#### Figure 3
**Electron energy transfer rates for the first seven excited states**. Plot of the present electron energy transfer rates (eVcm$^3$s$^{-1}$), as a function of electron temperature (K), for first seven excited electronic states in H$_2$O, as labelled on the plot. Also shown is the sum of the energy transfer rates for all 25 electronic states.

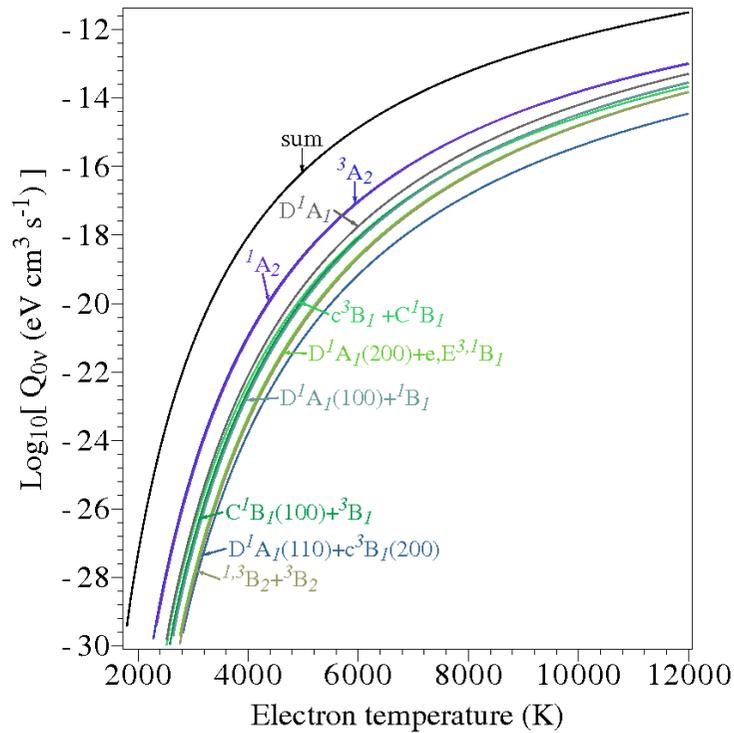

**Figure 4**
**Electron energy transfer rates for higher energy electronic states**. Plot of the present electron energy transfer rates (eVcm$^3$s$^{-1}$), as a function of electron temperature (K), for more of the excited electronic states in H$_2$O (states 10–18 of table 1 and as labelled in the plot). Also shown is the sum of the energy transfer rates for all 25 electronic states.

culations of the electron energy transfer rates for vibrational and electronic state excitations in H$_2$O are plotted in figures 3, 4, 5 and 6. This is the first time such computations have been attempted for these processes, and as such there are no other data against which we can compare the present results. Nonetheless it is clear from these plots that the energy transfer rates are considerably smaller for the electronic transitions (see figures 3, 4 and 5), compared to the vibrational transitions (see figure 6). Hence electronic excitations in H$_2$O represent a much smaller contribution to the electron cooling process than the vibrational excitations, when dealing with a thermal electron energy distribution.

To assist modellers who would use data such as these in their studies on upper atmosphere climatology, for example, we have fitted a power series function, via a least squares method, to each of the energy transfer rates in figures 3, 4, 5 and 6. That power series has the form given below in equation (6), and the resulting coefficients of the fitted power series are listed in table 2. Note that although not explicitly shown, the resulting fits using equation (6) to our respective energy transfer rates were all excellent. Indeed a reduced $\chi^2$ value [24] close to unity was achieved in each case.

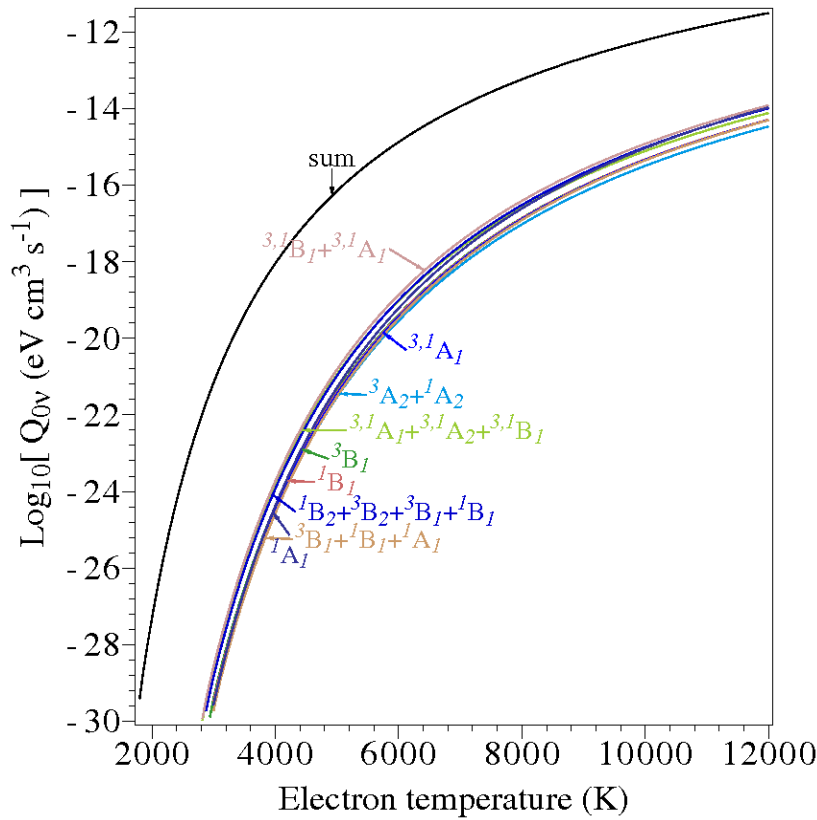

**Figure 5**
**Electron energy transfer rates for highest energy electronic states**. Plot of the present electron energy transfer rates (eVcm$^3$s$^{-1}$), as a function of electron temperature (K), for the remaining nine excited electronic states in H$_2$O (states 19–27 of table 1 and as labelled in the plot). Also shown is the sum of the energy transfer rates for all 25 electronic states.

$$log(Q_i) = A_i + B_i T_e^{1/2} + C_i T_e + D_i T_e^{3/2} + E_i T_e^2 + F_i T_e^{5/2} + G_i T_e^3 + H_i T_e^{7/2} + I_i T_e^4 \qquad (6)$$

The electron impact excitation rates, as calculated using equation (2), are plotted in figures 7 and 8. It is clear from these figures that, similar to the case for the energy transfer rates, the excitation rates for the vibrational transitions were much greater than those for the electronic states. This propensity was found to increase at the higher altitudes and was essentially caused by the very large difference in the flux of electrons with energy above the excitation thresholds for the vibrational states, compared to the corresponding flux of electrons with energies above the thresholds of the electronic states. As can be seen from figure 2, the flux of electrons with energy great enough to excite the vibrational states was ≈ 10 times greater at 80 km, and ≈ 10$^4$ times greater at 350 km, than the flux of electrons with energy above the excitation thresholds for the electronic states.

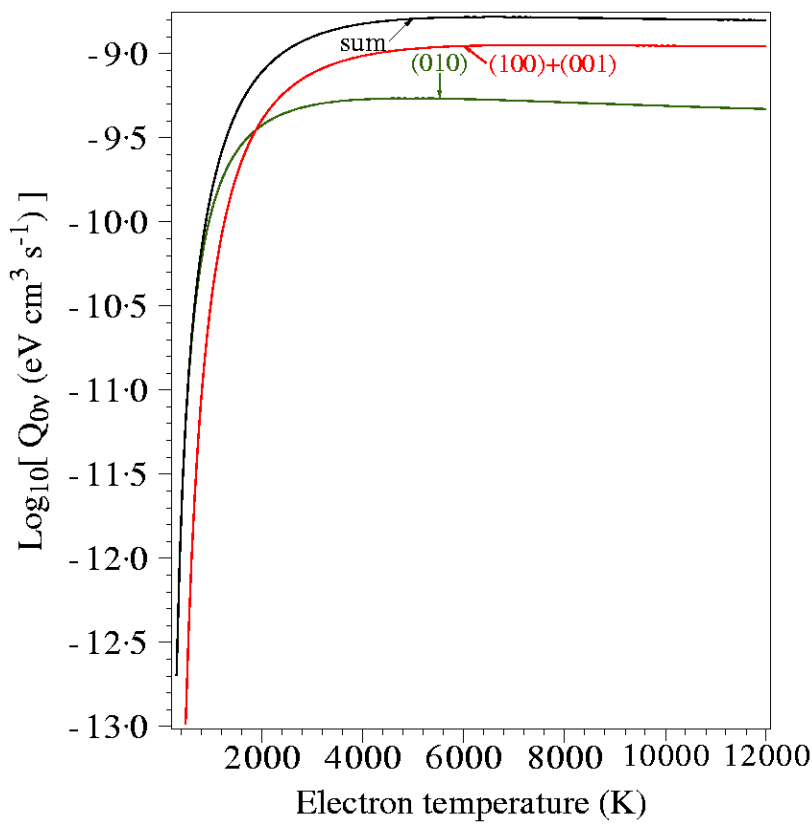

**Figure 6**
**Electron energy transfer rates for the vibrational levels**. Plot of the present electron energy transfer rates (eVcm$^3$s$^{-1}$), as a function of electron temperature (K), for the (010) and (100)+(001) vibrational levels of H$_2$O as labelled in the plot. Also shown is the sum of the energy transfer rates for these vibrational transitions.

Concentrating now solely on the electronic-state excitation rates, we find that for a given altitude there is a large variation in the calculated rates for the different electronic states. This simply reflects (see figure 1) the difference in the magnitude and energy dependence of the various integral cross sections for these states. It is also apparent from figures 7 and 8 that the qualitative (shape) form for the altitude dependence of each electronic state excitation rate is similar in each case. Namely, the excitation rate rises significantly in magnitude between altitudes of 80–130 kms before reaching a plateau at ≈ 150 kms. Thereafter (> 150 kms) there is only a more modest variation of the excitation rates with altitude. This behaviour can be understood by considering the secondary electron energy distributions in figure 2, where for altitudes between 130 km and 350 kms and for electron energies greater than ≈ 7 eV (see table 1) there is not a significant variation in the relevant energy distributions with altitude.

## 3. Conclusion

We have reported nineteen new integral cross sections for excitation of electronic states in water and described the use of these and our other recently measured cross sections for electronic state

**Table 2: Power series fitted to energy transfer rates**

| State | $A_i$ | $B_i$ | $C_i$ | $D_i$ | $E_i$ |
|---|---|---|---|---|---|
| (010) | -37.98 | 3.32 | $-1.76 \times 10^{-1}$ | $5.37 \times 10^{-3}$ | $-1.02 \times 10^{-4}$ |
| (100)+(001) | -53.94 | 4.72 | $-2.28 \times 10^{-1}$ | $6.44 \times 10^{-3}$ | $-1.14 \times 10^{-4}$ |
| $\tilde{a}^3 B_1$ | -192.48 | 7.40 | $-2.72 \times 10^{-2}$ | $-4.93 \times 10^{-3}$ | $1.50 \times 10^{-4}$ |
| $\tilde{A}^1 B_1$ | -182.22 | 5.46 | $9.05 \times 10^{-2}$ | $-8.62 \times 10^{-3}$ | $2.18 \times 10^{-4}$ |
| $^3 A_2$ | -217.11 | 6.49 | $1.05 \times 10^{-1}$ | $-9.87 \times 10^{-3}$ | $2.46 \times 10^{-4}$ |
| $^1 A_2$ | -175.83 | 3.05 | $2.18 \times 10^{-1}$ | $-1.17 \times 10^{-2}$ | $2.57 \times 10^{-4}$ |
| $\tilde{b}^3 A_1$ | -187.82 | 3.90 | $1.88 \times 10^{-1}$ | $-1.10 \times 10^{-2}$ | $2.47 \times 10^{-4}$ |
| $\tilde{B}^1 A_1$ | -205.30 | 4.71 | $1.88 \times 10^{-1}$ | $-1.19 \times 10^{-2}$ | $2.73 \times 10^{-4}$ |
| $\tilde{d}^3 A_1$ | -185.00 | 1.59 | $3.65 \times 10^{-1}$ | $-1.72 \times 10^{-2}$ | $3.68 \times 10^{-4}$ |
| $\tilde{c}^3 B_1 + \tilde{C}^1 B_1$ | -198.45 | 1.04 | $4.71 \times 10^{-1}$ | $-2.18 \times 10^{-2}$ | $4.71 \times 10^{-4}$ |
| $\tilde{D}^1 A_1$ | -190.37 | 1.68 | $3.75 \times 10^{-1}$ | $-1.77 \times 10^{-2}$ | $3.79 \times 10^{-4}$ |
| $\tilde{C}^1 B_1(100) + ^3 B_1$ | -252.84 | 6.34 | $2.37 \times 10^{-1}$ | $-1.58 \times 10^{-2}$ | $3.74 \times 10^{-4}$ |
| $^1 B_1 + \tilde{D}^1 A_1 (100)$ | -235.91 | 6.31 | $1.56 \times 10^{-1}$ | $-1.17 \times 10^{-2}$ | $2.76 \times 10^{-4}$ |
| $^3 A_2$ | -255.52 | 6.76 | $1.97 \times 10^{-1}$ | $-1.41 \times 10^{-2}$ | $3.35 \times 10^{-4}$ |
| $\tilde{D}^1 A_1(110) + \tilde{C}^1 B_1(200)$ | -232.55 | 6.21 | $1.36 \times 10^{-1}$ | $-1.04 \times 10^{-2}$ | $2.43 \times 10^{-4}$ |
| $^1 A_2$ | -218.73 | 3.60 | $3.04 \times 10^{-1}$ | $-1.58 \times 10^{-2}$ | $3.46 \times 10^{-4}$ |
| $\tilde{e}^3 B_1 + \tilde{E}^1 B_1 + \tilde{D}^1 A_1 (200)$ | -214.99 | 3.31 | $3.12 \times 10^{-1}$ | $-1.60 \times 10^{-2}$ | $3.47 \times 10^{-4}$ |
| $^3 B_2 + ^1 B_2 + ^3 B_2$ | -253.22 | 4.24 | $3.92 \times 10^{-1}$ | $-2.08 \times 10^{-2}$ | $4.64 \times 10^{-4}$ |
| $^3 A_1 + ^1 A_2 + ^1 A_1 + ^3 A_2 + ^1 B_1 + ^3 B_1$ | -340.25 | 12.15 | $7.66 \times 10^{-2}$ | $-1.36 \times 10^{-2}$ | $3.62 \times 10^{-4}$ |
| $^1 B_1 + ^3 B_1 + ^3 A_1 + ^1 A_1$ | -320.73 | 11.00 | $9.06 \times 10^{-2}$ | $-1.31 \times 10^{-2}$ | $3.40 \times 10^{-4}$ |
| $^1 B_2 + ^3 B_2 + ^3 B_1 + ^1 B_1$ | -250.50 | 5.83 | $2.29 \times 10^{-1}$ | $-1.43 \times 10^{-2}$ | $3.25 \times 10^{-4}$ |
| $^3 A_2 + ^1 A_2$ | -211.30 | 2.77 | $3.18 \times 10^{-1}$ | $-1.54 \times 10^{-2}$ | $3.24 \times 10^{-4}$ |
| $^3 B_1$ | -255.49 | 3.63 | $4.29 \times 10^{-1}$ | $-2.16 \times 10^{-2}$ | $4.71 \times 10^{-4}$ |
| $^1 B_1$ | -206.46 | 2.94 | $2.80 \times 10^{-1}$ | $-1.37 \times 10^{-2}$ | $2.86 \times 10^{-4}$ |
| $^3 A_1 + ^1 A_1$ | -218.13 | 3.36 | $2.91 \times 10^{-1}$ | $-1.46 \times 10^{-2}$ | $3.10 \times 10^{-4}$ |
| $^1 A_1 + ^3 B_1 + ^1 B_1$ | -244.26 | 4.12 | $3.34 \times 10^{-1}$ | $-1.75 \times 10^{-2}$ | $3.80 \times 10^{-4}$ |
| $^1 A_1$ | -217.71 | 3.18 | $3.01 \times 10^{-1}$ | $-1.49 \times 10^{-2}$ | $3.14 \times 10^{-4}$ |

**Table 2: Power series fitted to energy transfer rates** *(Continued)*

| State | $F_i$ | $G_i$ | $H_i$ | $I_i$ |
|---|---|---|---|---|
| (010) | $1.23 \times 10^{-6}$ | $-9.06 \times 10^{-9}$ | $3.75 \times 10^{-11}$ | $6.66 \times 10^{-14}$ |
| (100)+(001) | $1.29 \times 10^{-6}$ | $-9.02 \times 10^{-9}$ | $3.56 \times 10^{-11}$ | $-6.06 \times 10^{-14}$ |
| $\tilde{a}^3 B_1$ | $-2.12 \times 10^{-6}$ | $1.67 \times 10^{-8}$ | $-6.99 \times 10^{-11}$ | $1.23 \times 10^{-13}$ |
| $\tilde{A}^1 B_1$ | $-2.91 \times 10^{-6}$ | $2.22 \times 10^{-8}$ | $-9.16 \times 10^{-11}$ | $1.59 \times 10^{-13}$ |
| $^3 A_2$ | $-3.23 \times 10^{-6}$ | $2.42 \times 10^{-8}$ | $-9.82 \times 10^{-11}$ | $1.68 \times 10^{-13}$ |
| $^1 A_2$ | $-3.17 \times 10^{-6}$ | $2.28 \times 10^{-8}$ | $-8.96 \times 10^{-11}$ | $1.50 \times 10^{-13}$ |
| $\tilde{b}^3 A_1$ | $-3.06 \times 10^{-6}$ | $2.20 \times 10^{-8}$ | $-8.66 \times 10^{-11}$ | $1.44 \times 10^{-13}$ |
| $\tilde{B}^1 A_1$ | $-3.45 \times 10^{-6}$ | $2.52 \times 10^{-8}$ | $-1.00 \times 10^{-10}$ | $1.69 \times 10^{-13}$ |
| $\tilde{d}^3 A_1$ | $-4.49 \times 10^{-6}$ | $3.21 \times 10^{-8}$ | $-1.26 \times 10^{-10}$ | $2.10 \times 10^{-13}$ |
| $\tilde{c}^3 B_1 + \tilde{C}^1 B_1$ | $-5.80 \times 10^{-6}$ | $4.19 \times 10^{-8}$ | $-1.66 \times 10^{-10}$ | $2.79 \times 10^{-13}$ |
| $\tilde{D}^1 A_1$ | $-4.63 \times 10^{-6}$ | $3.31 \times 10^{-8}$ | $-1.30 \times 10^{-10}$ | $2.17 \times 10^{-13}$ |
| $\tilde{C}^1 B_1(100)+^3 B_1$ | $-4.80 \times 10^{-6}$ | $3.55 \times 10^{-8}$ | $-1.43 \times 10^{-10}$ | $2.43 \times 10^{-13}$ |
| $^1 B_1 + \tilde{D}^1 A_1(100)$ | $-3.50 \times 10^{-6}$ | $2.55 \times 10^{-8}$ | $-1.01 \times 10^{-10}$ | $1.69 \times 10^{-13}$ |
| $^3 A_2$ | $-4.28 \times 10^{-6}$ | $3.15 \times 10^{-8}$ | $-1.26 \times 10^{-10}$ | $2.12 \times 10^{-13}$ |
| $\tilde{D}^1 A_1(110)+ \tilde{C}^1 B_1(200)$ | $-3.04 \times 10^{-6}$ | $2.18 \times 10^{-8}$ | $-8.50 \times 10^{-11}$ | $1.40 \times 10^{-13}$ |
| $^1 A_2$ | $-4.23 \times 10^{-6}$ | $3.01 \times 10^{-8}$ | $-1.17 \times 10^{-10}$ | $1.93 \times 10^{-13}$ |
| $\tilde{e}^3 B_1 + \tilde{E}^1 B_1 + \tilde{D}^1 A_1(200)$ | $-4.24 \times 10^{-6}$ | $3.02 \times 10^{-8}$ | $-1.18 \times 10^{-10}$ | $1.94 \times 10^{-13}$ |
| $^3 B_2+^1 B_2+^3 B_2$ | $-5.76 \times 10^{-6}$ | $4.16 \times 10^{-8}$ | $-1.64 \times 10^{-10}$ | $2.74 \times 10^{-13}$ |
| $^3 A_1+^1 A_2+^1 A_1+^3 A_2+^1 B_1+^3 B_1$ | $-4.85 \times 10^{-6}$ | $3.65 \times 10^{-8}$ | $-1.48 \times 10^{-10}$ | $2.52 \times 10^{-13}$ |
| $^1 B_1+^3 B_1+^3 A_1+^1 A_1$ | $-4.51 \times 10^{-6}$ | $3.38 \times 10^{-8}$ | $-1.36 \times 10^{-10}$ | $2.31 \times 10^{-13}$ |
| $^1 B_2+^3 B_2+^3 B_1+^1 B_1$ | $-4.02 \times 10^{-6}$ | $2.88 \times 10^{-8}$ | $-1.12 \times 10^{-10}$ | $1.85 \times 10^{-13}$ |
| $^3 A_2+^1 A_2$ | $-3.86 \times 10^{-6}$ | $2.68 \times 10^{-8}$ | $-1.02 \times 10^{-10}$ | $1.66 \times 10^{-13}$ |
| $^3 B_1$ | $-5.75 \times 10^{-6}$ | $4.09 \times 10^{-8}$ | $-1.59 \times 10^{-10}$ | $2.63 \times 10^{-13}$ |
| $^1 B_1$ | $-3.37 \times 10^{-6}$ | $2.32 \times 10^{-8}$ | $-8.75 \times 10^{-11}$ | $1.40 \times 10^{-13}$ |
| $^3 A_1+^1 A_1$ | $-3.69 \times 10^{-6}$ | $2.56 \times 10^{-8}$ | $-9.75 \times 10^{-11}$ | $1.57 \times 10^{-13}$ |
| $^1 A_1+^3 B_1+^1 B_1$ | $-4.61 \times 10^{-6}$ | $3.26 \times 10^{-8}$ | $-1.26 \times 10^{-10}$ | $2.06 \times 10^{-13}$ |
| $^1 A_1$ | $-3.72 \times 10^{-6}$ | $2.58 \times 10^{-8}$ | $-9.82 \times 10^{-11}$ | $1.58 \times 10^{-13}$ |

Coefficients that result from the fitting of a power series to our computed electron energy transfer rates. The vibrational and electronic state transitions are labelled in the first column.

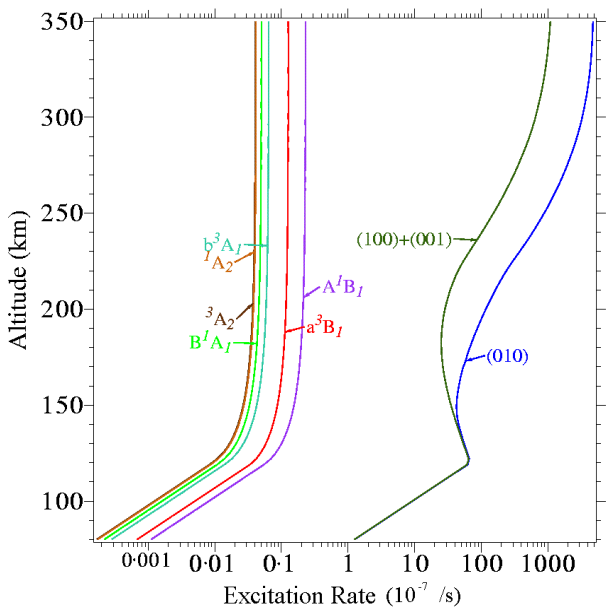

**Figure 7**
**Excitation Rates for lower energy transitions**. Present electron impact excitation rates ($10^{-7}$/s), as a function of altitude (km), for the vibrationally excited states and first six excited electronic states of $H_2O$ as labelled in the plot.

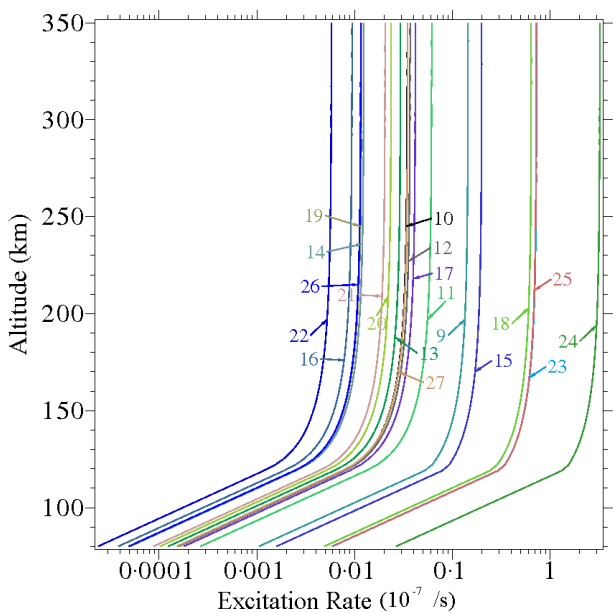

**Figure 8**
**Excitation Rates for higher energy transitions**. Present electron impact excitation rates ($10^{-7}$/s), as a function of altitude (km), for the remaining excited electronic states pertaining to this study. These remaining states are labelled 9 through 27, where 9 refers to the $\tilde{d}^3 A_1$ state, 27 refers to the final $^1A_1$ state and the rest are in the same order as listed in table 1.

excitation, as well as previously determined cross sections [13] for vibrational excitation, to calculate atmospherically relevant electron energy transfer rates and excitation rates for these inelastic processes in $H_2O$. It was found that vibrational excitation represents a more dominant process than electronic state excitation, when dealing with both a thermal electron distribution and an auroral secondary electron distribution. As this was the first time that these rates have been determined for these processes, there were no other data for us to compare with our results.

The calculation of these rates was an important first step towards including $H_2O$ in our statistical equilibrium code of the Earth's Ionosphere. However, more data are needed before $H_2O$ can be fully integrated into the statistical equilibrium code including transition probabilities, quenching rates and atmospheric densities of $H_2O$ at the relevant altitudes. Hence no new conclusions can yet be drawn about the origins of the Mienel bands. However, once these extra data are determined, or successfully located in the existing literature, the statistical equilibrium code will be updated and a full investigation for electron-driven processes in atmospheric $H_2O$ will be conducted.

### Acknowledgements
This work was supported in part by the Australian Research Council through its Centres of Excellence Programme. One of us (PT) also wishes to thank the Ferry Trust for financial support.

### References
1.　Dalgarno A: *Can J Chem* 1969, **47:**1723-1731.
2.　Vallance Jones A: *Aurora* Dordrecht, Holland: D Reidel; 1974.
3.　Meinel AB: *Astrophys J* 1950, **111:**555-564.
4.　Meinel AB: *Astrophys J* 1950, **112:**120-130.
5.　Cartwright DC: *J Geophys Res* 1978, **83:**517-531.
6.　Campbell L, Cartwright DC, Brunger MJ, Teubner PJO: *J Geophys Res* 2006, **111:**A09317.
7.　Jones DB, Campbell L, Bottema MJ, Teubner PJO, Cartwright DC, Newell WR, Brunger MJ: *Planet Space Sci* 2006, **54:**45-59.
8.　Cartwright DC, Brunger MJ, Campbell L, Mojarrabi B, Teubner PJO: *J Geophys Res* 2000, **105:**20857-20867.
9.　Harb T, Kedzierski W, McConkey JW: *J Chem Phys* 2001, **115:**5507-5511.
10.　Thorn PA, Brunger MJ, Teubner PJO, Diakomichalis N, Maddern T, Bolorizadeh MA, Newell WR, Kato H, Hoshino M, Tanaka H, Cho H, Kim YK: *J Chem Phys* 2007, **126:**064306.
11.　Thorn PA, Brunger MJ, Kato H, Hoshino M, Tanaka H: *J Phys B: At Mol Opt Phys* 2007, **40:**697-708.
12.　Brunger MJ, Thorn PA, Campbell L, Diakomichalis N, Kato H, Kawahara H, Hoshino M, Tanaka H, Kim YK: *Int J Mass Spectrom* 2008, **271:**80-84.
13.　Yousfi M, Benabdessadok MD: *J Appl Phys* 1996, **80:**6619-6630.
14.　Thorn PA: **Electronic State Excitations in the Water Molecule by Collisions with Low Energy Electrons.** In *PhD thesis* School of Chemistry, Physics and Earth Sciences, Flinders University; 2008.
15.　Jones DB, Campbell L, Bottema MJ, Brunger MJ: *New J Phys* 2003, **5:**114.1-114.11.
16.　Pavlov AV: *Ann Geophys* 1998, **16:**176-182.
17.　Cho H, Park YS, Tanaka H, Buckman SJ: *J Phys B: At Mol Opt Phys* 2004, **37:**625-634.
18.　Campbell L, Brunger MJ, Nolan AM, Kelly LJ, Wedding AB, Harrison J, Teubner PJO, Cartwright DC, McLaughlin B: *J Phys B: At Mol Opt Phys* 2001, **34:**1185-1199.

19. Itikawa Y, Mason NJ: *J Phys Chem Ref Data* 2005, **34:**1-22.
20. **International Reference Ionosphere – IRI-2001**   [http://ccmc.gsfc.nasa.gov/modelweb/models/iri.html]
21. Lummerzheim D, Lilensten J: *Ann Geophys* 1994, **12:**1039-1051.
22. Feldman PD, Doering JP: *J Geophys Res* 1975, **80:**2808-2812.
23. Lummerzheim D, Rees MH, Anderson HR: *Planet Space Sci* 1989, **37:**109-129.
24. Bevington PR: *Data Reduction and Error Analysis for the Physical Sciences* New York: McGraw-Hill; 1969.